\documentclass[12pt]{iopart}

\usepackage{graphicx}
\begin{document}

\title{Limits to the energy resolution of a single Air Cherenkov Telescope at low energies}

\author{Dorota Sobczy\'{n}ska}

\address{University of {\L }\'{o}d\'{z}, Experimental Physics 
Department, Pomorska 149/153, 90-236 {\L }\'{o}d\'{z},Poland}
\ead{ds@kfd2.phys.uni.lodz.pl}
\begin{abstract}
The photon density on the ground is a fundamental quantity in all experiments based on Cherenkov light measurements, e.g.\ in the Imaging Air Cherenkov Telescopes (IACT). IACT's are commonly and successfully used in order to search and study Very High Energy (VHE) $\gamma$-ray sources. Difficulties with separating primary photons from primary hadrons (mostly protons) in Cherenkov experiments become larger at lower energies. I have calculated longitudinal and lateral density distributions and their fluctuations at low energies basing on Monte Carlo simulations (for vertical $\gamma$ cascades and protonic showers) to check the influence of the detector parameters on the possible measurement. Relative density fluctuations are significantly higher in proton than in photon induced showers. Taking into account the limited detector field of view (FOV) implies the changes of these calculated distributions for both types of primary particles and causes an enlargement in relative fluctuations. Absorption due to Rayleigh and Mie scattering has an impact on mean values but does not change relative fluctuations. The total number of Cherenkov photons is more sensitive to the observation height in $\gamma$ cascades than in proton showers at low primary energies. The relative fluctuations of the density do not depend on the reflector size in the investigated size range (from 240 $m^2$ up to 960 $m^2$). This implies that a single telescope with a mirror area larger than that of the MAGIC telescope cannot achieve better energy resolution than estimated and presented in this paper. 
The correlations between longitudinal and lateral distributions are much more pronounced for primary $\gamma$-ray than for primary proton showers.

\end{abstract}

\pacs{95.55.Ka;95.55.Vj;95.75-z;95.85.Pw;95.85.Ry}
\noindent{\it Keywords}: VHE $\gamma$-astronomy, Extensive air shower, Cherenkov photon density
\submitto{\JPG}
\maketitle

\section{Introduction}

Ground-based $\gamma$-astronomy has developed very fast since the discovery of the first TeV $\gamma$-ray source, the Crab Nebula in 1989 \cite{whipple} by the Whipple collaboration. 
This experiment was based on the imaging air Cherenkov technique, which is still the best method of $\gamma$-ray measurement on the ground. The main idea of this technique is the collection of Cherenkov light (produced in the atmosphere by very energetic charged particles from the shower) using a telescope. The light is recorded by a camera (matrix of photo multipliers), which is mounted in the focal plane of the telescope. Light shower images obtained by this measurement are finally analysed to select primary $\gamma$-ray candidates from the several orders of magnitude larger hadronic background. The method of gamma/hadron separation proposed by Hillas in 1985 \cite{hillas} is commonly used.

From the experimental point of view, it is very important  how many Cherenkov photons are collected by the telescope. Events containing not enough light cannot trigger the telescope and thus cannot be registered. The larger the area of the telescope mirror, the more Cherenkov photons from the same shower can be focused onto the camera. The expected image SIZE  (sum of the detected image amplitude) depends also on the telescope field of view and the altitude of the telescope \cite{aha97,portocar98}. 
The primary energy reconstruction is based on the measured image SIZE and the DIST parameter (DIST is the distance between the camera and image centres) or on the measured image SIZE and the reconstructed distance of the shower core to the telescope (impact paramter)\cite{aha99,hoffman00,unfol,hanna08}.\\
The Cherenkov photon density on the ground ($\rho$) is (due to all those reasons) a physical quantity, which is especially worth investigating. Several publications have focussed on this subject. Fluctuations of the Cherenkov light density on the ground have been studied in \cite{portocar98,bhat98,sinha95}, but were calculated for smaller telescopes than the currently used and planned generation. The possible $\gamma$/hadron separation using the Cherenkov light density fluctuations is presented in \cite{bhat02,contreras06}.

Fluctuations of the first interaction height cause fluctuations of the total number of produced Cherenkov photons (I shall call it the shower size) at a fixed energy.
In the case of primary $\gamma$-rays the shower size depends on the height of the first $e^+$ $e^-$  pair production ($h_{0}$): 
the smaller the height of the first pair production, the larger the shower size \cite {bhat98}.
The density of the atmosphere and the refraction index are diminishing functions of the height and therefore the energy threshold for Cherenkov light production is lower in the lower atmosphere. When the cascade starts deeper, more $e^+$ and $e^-$ have an energy which is above those thresholds and produce Cherenkov light. This results in a larger shower size.
In the case of primary protons the correlations between  the shower size and the first interaction height are not observed \cite {bhat98}.

In this paper the longitudinal development and expected total number of Cherenkov photons of proton induced showers and electromagnetic cascades are calculated for photons with unlimited and limited zenith angles.\\
The influence of the telescope parameters (like area, FOV and altitude) on the possible measurment (by an ideal single IACT) of the Cherenkov photon density on the ground and its fluctuations is shown and discussed. A detector area of 240.25 $m^2$ has been chosen to show the density fluctuations obtained for the largest working Cherenkov telescope - MAGIC \cite{bario,bax04,albert}. The density has been calculated from Cherenkov photons which may hit the camera of the telescope in order to estimate the expected relative fluctuations of the image SIZE for very large IACT. Similar fluctuations of the Cherenkov light density are obtained and they are presented for even larger detectors - 480 and 960 $m^2$ to check if enlargement of the mirror results in an improvement of the energy resolution. The Cherenkov light density fluctuations have not been presented for such large detectors.\\   
Additionally, the correlations between the Cherenkov light produced at a fixed depth in the atmosphere and the number of photons which hit the camera  of the single telescope on the ground are calculated and shown. All results are obtained by Monte Carlo simulations.

\section{Monte Carlo simulations}
The simulations in this paper used the CORSIKA code version 6.023 \cite{heck,knapp} with GHEISHA and VENUS as low and high energy (primary momentum above 80 GeV/c) interaction models for the primary protons. I have modified the CORSIKA code slightly in order to analyse the simulated Cherenkov light without writing the informations of each individual photon to the output file. All simulations were done using the US standard atmosphere model. It is expected (and I have checked) that the characteristics of the produced Cherenkov photons depend on the chosen atmosphere model, but this is not investigated in this paper. 
The MC simulations have been performed for the MAGIC site \cite{bario,bax04,albert}, that is 2200 m above sea level (around 800 $g/cm^2$). 
For the vertical $\gamma$- cascades fixed primary energies of 20, 50, 100, 200, 300 and 500 GeV have been simulated. Primary energies of 50, 100, 200, 500 and 1000 GeV have been chosen for vertical proton induced showers respectively. The impact parameter of the showers has been fixed to the position x=0, y=0 on the ground. The square area of 900 m * 900 m was completely covered by square detectors of 15.5 m *15.5 m (240.25 $m^2 $). The numbers of produced Cherenkov photons (wavelength between 290 and 600 nm) hitting the detectors in each shower have been used to calculate the Cherenkov photon densities as well as their fluctuations. This configuration of the detectors also enabled me to obtain  the expected densities and fluctuations for two and four times larger detectors. 
The Cherenkov photons with zenith angle below 2.5 $^o$ have been counted in the MC set II of simulations because real Cherenkov telescopes have a limited FOV. 
Additionally, the effect of light absorption in the atmosphere (Rayleigh and Mie scattering according to the Sokolsky formula \cite{sokol}) was taken into account in the MC set III of the simulations. In this MC set photons with zenith angle belov  2.5 $^o$ which were not absorbed in the atmosphere were analysed (I shall call them photons seen by the telescope).   
Taking into account both effects (limited FOV of the telescope and light absorption by the atmosphere) allowed me to show the expected fluctuations in the real experiments in a more realistic way. 
The last MC set (IV) of the simulations was performed for the observation level altitude of 4 km a.s.l. (with the same geomagnetic field) in order to check how the detector location height influences the capability of the Cherenkov light measurment.  
Overviews of parameters and the number of simulated events in all MC sets are presented in Table 1 and Table 2 respectively.\\

\begin{table}
\caption {\label{tab1}Parameters of MC sets}
\begin{indented}
\item[]\begin{tabular}{@{}*{5}{l}}
\br
 MC set   & I & II & III & IV\cr
\mr
Observation altitude [km]& 2.2 & 2.2 & 2.2 & 4.0\cr
Telescope's FOV [deg]    & -   & 5.0 & 5.0 & 5.0\cr
Atmosperic absorption    & No  & No  & Yes & Yes\cr
\br
\end{tabular}
\end{indented}
\end{table}

\begin{table}
\caption {\label{tab2}Number of simulated events in all MC sets (see the text)}
\begin{indented}
\item[]\begin{tabular}{@{}*{6}{l}}
\br
primary particle & primary energy  & MC I & MC II & MC III & MC IV\cr
\mr
$\gamma$ &  20 GeV & 40000 & 40000 & 20000 & 20000\cr
$\gamma$ &  50 GeV & 40000 & 40000 & 20000 & 20000\cr
$\gamma$ & 100 GeV & 40000 & 40000 & 20000 & 20000\cr
$\gamma$ & 200 GeV & 20000 & 20000 & 10000 & 10000\cr
$\gamma$ & 300 GeV & 10000 & 10000 & 10000 & 10000\cr
$\gamma$ & 500 GeV & 5000 & 5000 & 5000 & 5000\cr
proton &  50 GeV & 40000 & 40000 & 40000 & 40000\cr
proton & 100 GeV & 40000 & 40000 & 40000 & 40000\cr
proton & 200 GeV & 20000 & 20000 & 20000 & 20000\cr
proton & 500 GeV & 20000 & 20000 & 20000 & 20000\cr
proton & 1000 GeV & 10000 & 10000 & 10000 & 10000\cr

\br
\end{tabular}
\end{indented}
\end{table}
\section{Results and discussion}

\subsection{Longitudinal distribution}
Figure 1a illustrates the dependence between the shower size ($N_{t}$) and the height of the first $e^+$ $e^-$  pair production ($h_{0}$)
for a primary $\gamma$ energy of 50 GeV. This figure has been obtained from the MC set I. As expected, the total number of the Cherenkov photons produced in the shower is larger for electromagnetic cascades starting deeper in the atmosphere. Similar characteristics are observed for all energies of the primary $\gamma$-ray. 
The correlations between the shower size and the first interaction height are not observed in proton showers, which is presented in figure 1b. As an example, a primary energy of 100 GeV (MC set I) has been chosen. Similar correlations between the shower size and the height of the first interaction were presented in \cite {bhat98}.

\begin{figure}
\begin{center}
\includegraphics*[width=14cm]{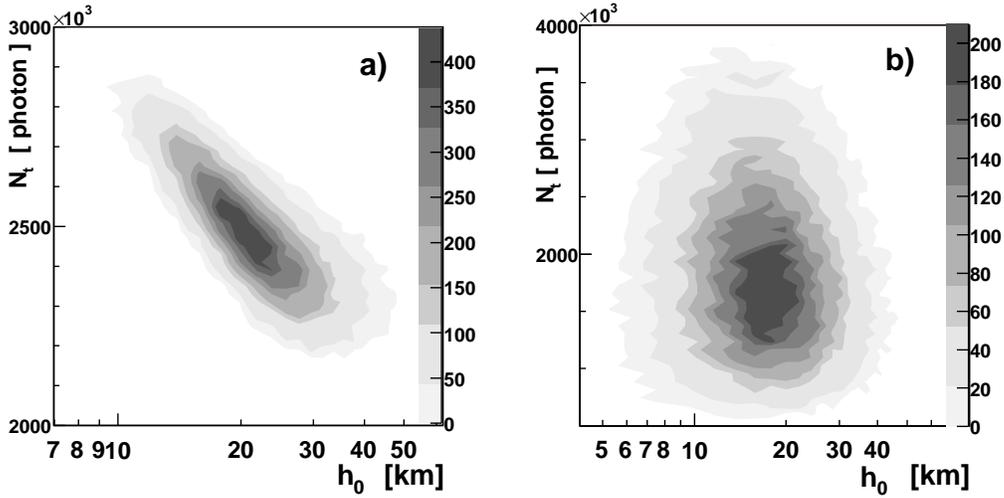}
\end{center}
\caption{The total number of produced Cherenkov photons versus the first interaction altitude: {\bf a)} $\gamma$ cascade (50 GeV); {\bf b)} proton induced shower (100 GeV). Both figures show the results of the MC set I. The grey colour scale denotes the number of MC events.}

\label{total_size_heigh}
\end{figure} 

The depth of the first interaction fluctuates which causes fluctuations in the distribution of the Cherenkov photon production height (so called longitudinal distribution $n_{ph}= dN_{t} / dt$).   
Figure 2a presents the average number of Cherenkov photons generated in a thickness of 1 $g/cm^2$ versus the atmospheric depth for the primary $\gamma$-ray. The MC simulations set III has been chosen because it contains the conditions of the real experiment (FOV and the atmosperic absorption). Most of Cherenkov photons in the $\gamma$ cascades come from depths of above 600 $g/cm^2$ at all simulated primary energies. The comparison between different MC sets (I, II and III) is shown in figure 2c for a primary energy of 20 GeV and 500 GeV. The impacts of the limited zenith angle for the Cherenkov photons and the light absorption on the calculated longitudinal distribution are similar for the lowest and the highest simulated primary energies. 

The position of the average cascade maximum for photons obtained from MC set III (solid line in the figure) is around 20 $g/cm^2$ lower than for all produced photons (dotted line in the plot) in the whole investigated energy range. These differences are caused mainly by the limited FOV of the telescope. The angular distribution of the generated Cherenkov photons is wider deeper in the atmosphere due to two reasons. The first of them is the larger Cherenkov angle because of the higher refraction index. The second effect is the wider angular distribution of $e^+$ and $e^-$ because deeper in the atmosphere their average energy is lower, which results in a larger angle after the multiscattering effect. 
The influence of the atmospheric absorption is opposite. The probability of light absorption in the atmosphere increases as the production height increases. It has been verified that the RMS deviation of the shower maximum remains stable on the level of 70 $g/cm^2$ for all primary $\gamma$-ray MC simulation sets and all energies.\\
Figures 2b and 2d show the average longitudinal distributions for proton showers. The distributions are much flatter above the shower maximum than in $\gamma$ cascades because the longitudinal distribution of the charged particles is also flatter in proton induced showers than in electromagnetic cascades.
The fraction of photons with zenith angles above the detector limit is much larger at the lowest energy than at the highest simulated proton energy (compare the dotted and dashed lines in figure 2d). The depth of the average shower maximum for photons obtained in MC set III is around 20 $g/cm^2$ lower than for all produced photons (dotted line in the plot) for energies of 500 GeV and 1 TeV only. At lower energies there are smaller differences in the mean shower maximum position obtained from MC sets I and III. The RMS deviation of the shower maximum of all produced photons decreases from 140 $g/cm^2$ at 50 GeV to 130 $g/cm^2$ at 1 TeV. These fluctuations are enlarged by the inclusion of the limited FOV effect in the simulations at energies below 500 GeV. Finally, the dispersions of the shower maximum obtained from MC set III are 180 $g/cm^2$ at an energy of 50 GeV and 130 $g/cm^2$ at 1 TeV. 

The ratio of the RMS deviation to the average number of Cherenkov photons generated in 10 $g/cm^2$ ($RF_{n_{ph}}$) as a function of atmospheric depth for photons seen by the telescope (simulation set III) is presented in figure 3a and 3b for primary $\gamma$ and proton showers, respectively.
These relative fluctuations decrease with energy for depths larger than the position of the maximum of the $\gamma$ cascade, while higher in the atmosphere they almost do not depend on the energy. 
The higher the primary energy, the lower the ratio that has been obtained in proton induced showers for all atmospheric depths, except the highest simulated energies of 500 GeV and 1 TeV which do not differ below 200 $g/cm^2$ and above 700 $g/cm^2$.

The influence of the telescope FOV and atmospheric absorption on these calculated relative fluctuations is shown in figure 3c ($\gamma$-rays) and figure 3d (protons). 
Taking into account the limited FOV enlarges these relative dispersions, while the atmospheric absorption does not have any noticeable influence on them for both types of the primary particles (solid covered dashed lines).
The mean number of photons produced at the shower maximum presented in this paper are much higher than shown in \cite {bhat98}. The different wavelength range of the Cherenkov photons does not explain this discrepancy. The integration of the production height distribution from \cite {bhat98} gives a much smaller value than the presented shower size in the same paper.
The relative fluctuations shown in this section are comparable with those presented in \cite {bhat98}.

\begin{figure}
\begin{center}
\includegraphics*[width=14cm]{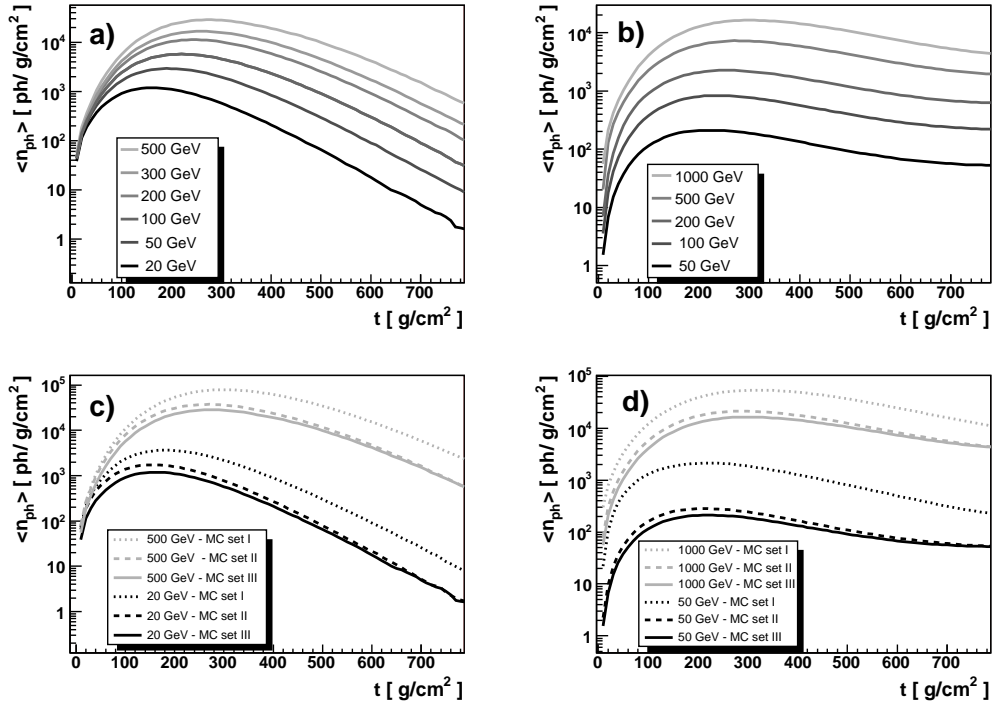}
\end{center}
\caption{ The number of produced Cherenkov photons versus atmospheric depth: {\bf a)} photons incident on the telescope mirror taking into accout inefficiency from atmospheric absorption and limited FOV (MC set III) - $\gamma$-rays; {\bf b)} the same like a) but for protons; {\bf c)} the influence of the limited FOV and atmospheric absorption on the possible measurment of the longitudinal development of $\gamma$-rays (dotted, dashed and solid lines correspond to the MC set I, II and III respectively); {\bf d)} the same as c), but for protons.}
 
\label{longitudinal}
\end{figure} 

\begin{figure}
\begin{center}
\includegraphics*[width=14cm]{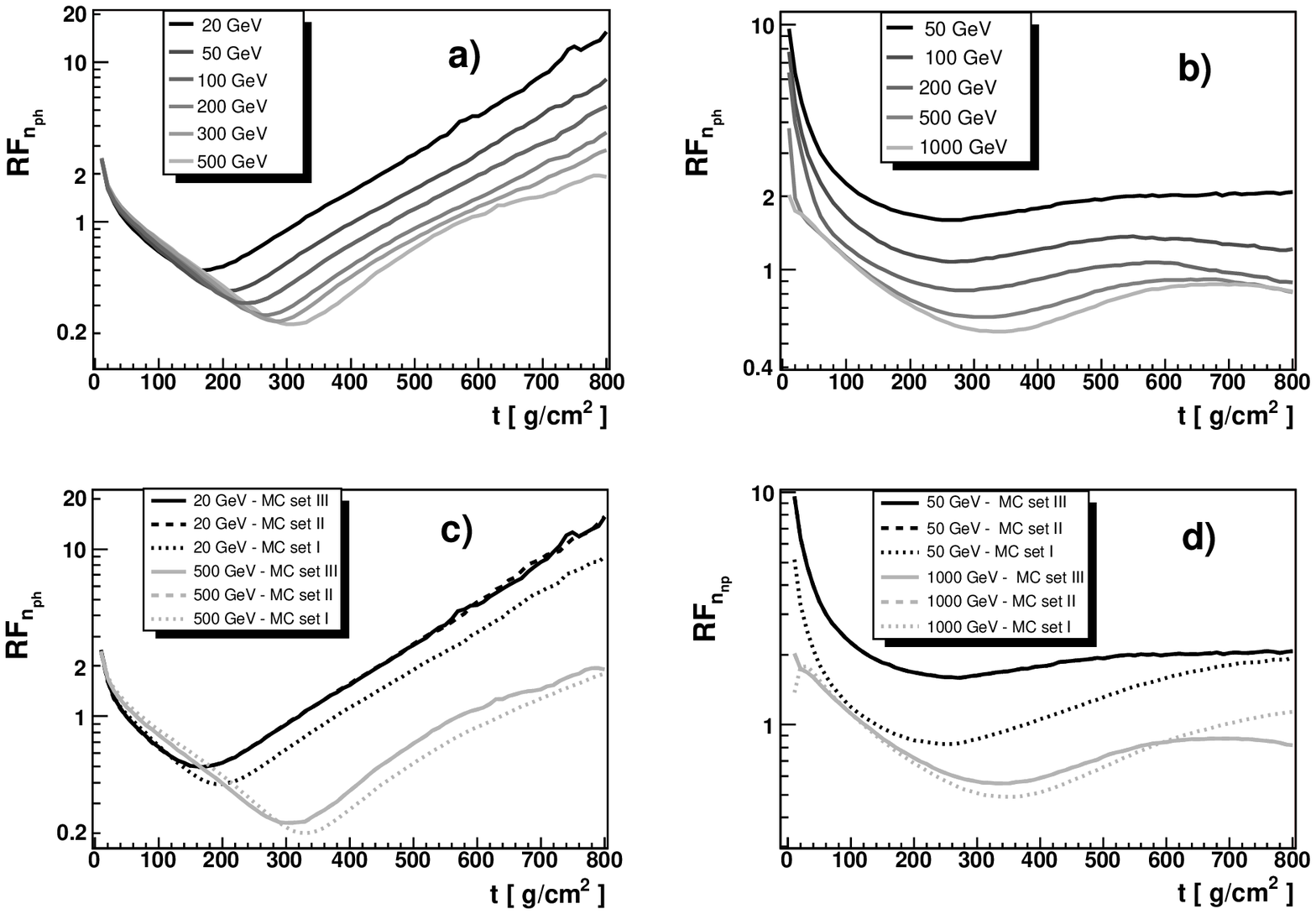}
\end{center}
\caption{Relative fluctuations of the number of produced photons versus depth of the atmosphere: {\bf a)} comparison of all simulated energies of the primary $\gamma$-rays (MC set III); {\bf b)} comparison of all simulated energies of the primary protons (MC set III); {\bf c)} the calculated $RF_{n_{ph}}$ for $\gamma$-rays: dotted lines correspond to all produced photons(MC set I); dashed lines correspond to all Cherenkov photons hitting the telescope mirror within the 5 deg. field of view, neglecting atmospheric absorption (MC set II);  solid lines correspond to  all photons incident on the telescope mirror taking into accout inefficiency from atmospheric absorption and limited FOV (MC set III); {\bf d)} the same as c), but for protons.}

\label{longitudinal_rel}
\end{figure}

\subsection{Shower size fluctuations}

The average total number of the Cherenkov photons (defined in introduction as the shower size) depends on the primary energy and primary particle type. Figures 4a and 4b show this for primary $\gamma$-rays and protons respectively. Results of all simulation sets are presented. The average shower size is a linear function of energy in all MC sets and both simulated primary particles as shown in \cite{bhat98}. The results of the $\gamma$ MC set III are consistent with those presented in \cite{portocar98} while I take into account that in \cite{portocar98} they were calculated from photons having zenith angles smaller than 1.4 $^o$ and are within a radius of 150 m. These two conditions may explain the non-linear dependence and different influence of the observation altitude on the shower sizes, which are presented in \cite{portocar98} for primary protons. As can be seen in Figure 4b the expected total number of Cherenkov photons, which were not absorbed and have zenith angles belov 2.5 $^o$, is aproximately the same on both simulated observation levels.    
 
The ratios of the RMS deviation of the shower size to the mean (relative fluctuations denoted as $RF_{N_{t}}$) are presented in figure 4c for $\gamma$ cascades. 
These relative dispersions decrease significantly with energy below 100 GeV, while for higher energies a weaker dependence on the energy has been obtained. This ratio calculated from MC set I decreases from $6\%$ to $5\%$, while the energy increases from 100 GeV to 500 GeV. 
Slightly lower results (from $5\%$ at 100 GeV to $3\%$ at 1 TeV) were published in \cite{bhat98}, but they were obtained from lower statistics and for different wavelengths, so the results are consistent. The limited FOV has significantly influenced the calculated ratio only for primary energies below 100 GeV (dashed line on the plot). 
The effect of the Cherenkov photon absorption in the atmosphere enlarges the relative fluctuations of the shower size at 2.2 km (solid line in figures) compared to 4 km (dashed-dotted line in figures).

The same quantity is shown in figure 4d for the proton primaries. The relative fluctuations of the shower size decrease from $40\%$ at 50 GeV to $29\%$ at 1 TeV in the case of MC set I and this result is also a little higher than that presented in \cite{bhat98}. The influences of the limited detector FOV and the atmospheric absorption are also presented in this figure. Both change significantly the relative dispersions of the shower size at energies below 300 GeV. The higher the observation level, the higher the ratios that have been obtained (inversely to the behaviour of $\gamma$ cascades).
This difference may be explained by a different longitudinal distribution (see section 3.1). For primary $\gamma$-rays most of the Cherenkov light is produced 4 km a.s.l. (approximately 600 $g/cm^2$) while for the primary protons the Cherenkov photons created lower in the atmosphere account for a non negligible amount of the shower size. 
 
The impact of the limited FOV of the detector and the height of the observation level on the calculated mean shower size is presented in figure 5a and 5b for primary $\gamma$ and proton showers respectively. The $RF_{N_{t}}$ is a ratio between the mean shower size obtained in MC sets II, III and IV to the average shower size calculated from the MC set I. The mean number of the photons in the limited FOV (dashed lines in plots) account for ca.\ $45\%$ of the shower size calculated from all the produced light in electromagnetic cascades in almost the whole simulated energy range. The effect of the light absorption reduces the expected shower size by an additional $7\%$ for the observation level of 4 km and $10\%$ for 2.2 km. In the case of primary $\gamma$-rays, the calculated ratio increases with energy for a primary energy below 100 GeV, at a higher energy it is constant.  In the case of primary protons the ratio between the average total number of photons fulfilling the detector criteria to the mean total shower size increases with energy. The limited detector FOV eliminates on average $85\%$ of the shower size calculated from all produced Cherenkov photons at 50 GeV. This value decreases to $62\%$ at 1 TeV. The reason of that can be seen in figure 2d: the fraction of photons with zenith angles outside the detector FOV is much larger at the lowest energy than at the highest simulated proton energy even much below the shower maximum. This is caused by differences in the zenith angle distribution of the secondary particles, which are produced in the hadronic interactions. For lower energy protons a wider distribution is expected and finally less produced photons are within the FOV. The average shower size at 2.2 km (soild line in the plot) is larger than at 4 km a.s.l. (dotted line in the plot). These results show that the mean number of absorbed Cherenkov photons between the two levels is smaller than the average number of produced photons between the two simulated observation altitudes.

\begin{figure}
\begin{center}
\includegraphics*[width=14cm]{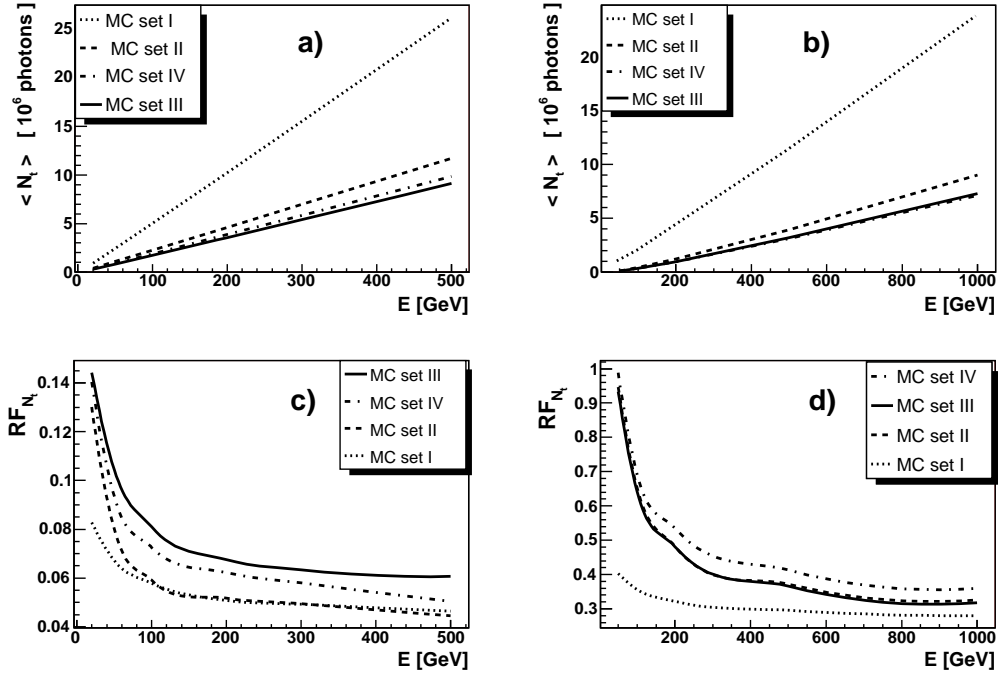}
\end{center}
\caption{{\bf a)} The average shower size in $\gamma$ cascades versus the primary energy. {\bf b)} The average shower size in proton showers versus the primary energy. {\bf c)} The relative fluctuations of the shower size for primary $\gamma$-rays. {\bf d)} The relative fluctuations of the shower size for proton primaries.}

\label{total size}
\end{figure} 
\begin{figure}
\begin{center}
\includegraphics*[width=14cm]{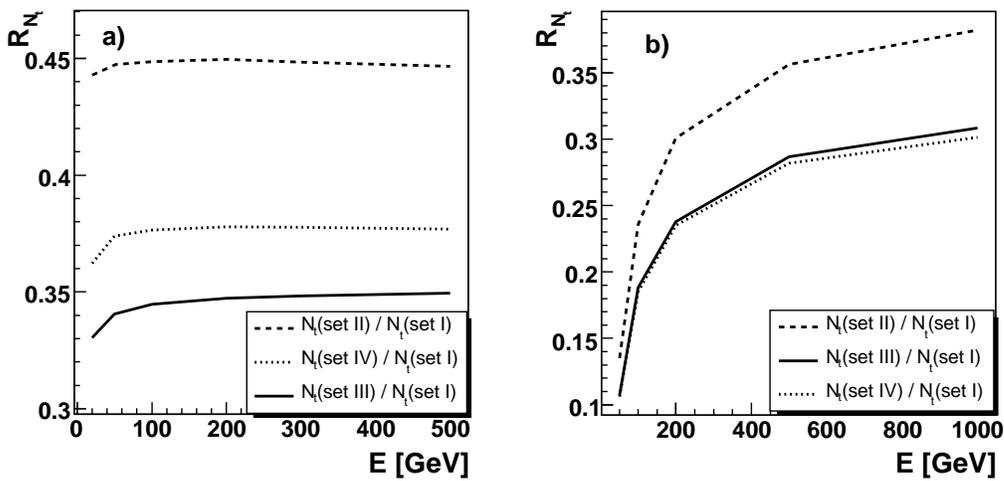}
\end{center}
\caption{The ratio of the average shower sizes calculated from MC sets: II, III and IV with respect to the mean total number of all produced photons as a function of the primary energy: {\bf a)} $\gamma$ cascades; {\bf b)} proton showers.}

\label{fraction}
\end{figure}

\subsection{The lateral density distribution of Cherenkov photons}

The average density of Cherenkov photons on the ground versus the distance from the core axis (so called average lateral density distribution) has been calculated for all detector positions (see MC section). Primary $\gamma$-rays at an energy of 50 GeV (MC set I) have been chosen as an example and are shown in figure 6b. The distribution is not symmetrical because the Earth's magnetic field influences the shower development. This effect is especially pronounced at low energies (50 GeV in this case), but has often been missed in higher energy (e.g. 1 TeV) simulations. Figure 6a has been obtained for the same primary energy but with very low magnitude of the Earth's magnetic field, and the asymmetry has vanished. 
The largest differences in lateral distributions are expected between the North-South (N-S) and East-West (E-W) directions. 
The average lateral distribution for both N-S (dashed lines in the plot) and E-W (dotted line in the plot) directions are demonstrated in figure 6c. The mean density has been calculated from all produced photons, but similar differences occur in all MC sets for primary $\gamma$-rays. 
The black solid line in the figure 6c shows the mean lateral density calculated from MC without the Earth's magnetic field. 
The highest density occurs at the impact parameter of around 120 m (so called hump \cite{raosinha95}).
The presence of the magnetic field causes a shift of electrons and positrons towards west and east directions, respectively. In the first approximation, both charges should create symmetric lateral distributions, but as they are shifted, their superposition cannot be symmetric. As a result, the hump is more pronounced in the N-S direction. The density decreases faster with the impact parameter in N-S than in E-W direction for impact parameters above the hump position. 
Asymmetry between N-S and E-W directions has not been found for showers initiated by protons.
For proton showers, the asymmetry still exists in the electromagnetic sub-showers, but it is washed out since a typical hadronic interaction produces a number of pions at large transverse momentum, and contains a number of electromagnetic sub-showers as well as contributions from single muons that are less deflected by the magnetic field. The influence of the Earth's magnetic field on Cherenkov light distibutions has been studied in \cite{porter73,bowden92}. The analysis of geomagnetic field effects on IACT's is presented in \cite{lang94,comm08}.
All the results presented below were calculated for the N-S direction.\\
\begin{figure}
\begin{center}
\includegraphics*[width=14cm]{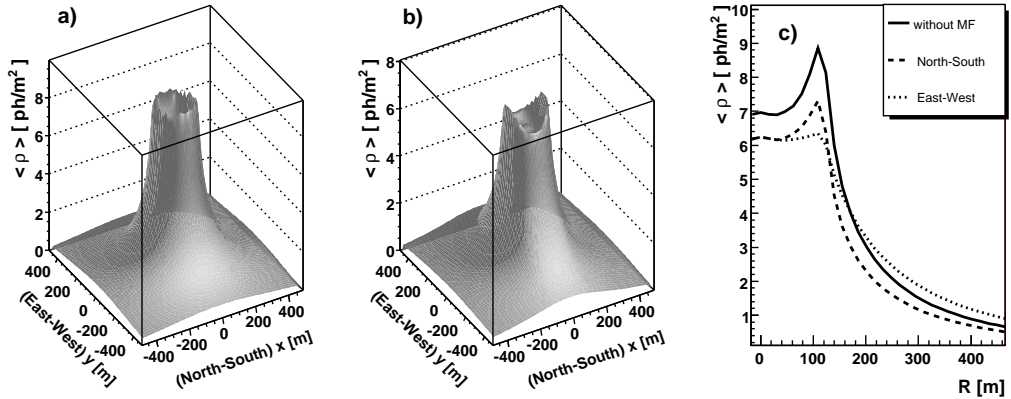}
\end{center}
\caption{The mean density of Cherenkov light for all produced photons in $\gamma$ cascades (50 GeV): {\bf a)} without magnetic field; {\bf b)} with magnetic field; {\bf c)} Comparison of the mean lateral density distributions obtained in North-South and East-West directions.}

\label{later_MF}
\end{figure} 
The impact of the telescope FOV and the light absorption on the calculated lateral distribution is shown in figures 7a and 7b for primary $\gamma$-rays and protons, respectively. The dotted lines correspond to the mean density calculated from MC set I, while the dashed lines represent the Cherenkov light in the camera FOV (MC set II). The solid lines are the results obtained for non-absorbed photons in the detector FOV (MC set III).
The highest (500 GeV for $\gamma$'s and 1 TeV for protons) and the lowest (20 GeV for $\gamma$'s and 50 GeV for protons) energies have been chosen to illustrate both effects, but all the densities obtained from MC set III are presented in the figures. 

The maximum of the average density is expected at the shower core, and the hump is less pronounced for $\gamma$'s with energies larger than 100 GeV.
Photons around the shower core position are produced close to the observation level if there are still energetic charged particles in the shower. This is therefore more likely for showers initiated by higher energy primaries that tend to have more particles reaching ground level. The hump is made by photons coming from depths around the shower maximum. 
For $\gamma$ cascades, the influence of the limited FOV on the calculated average lateral density distribution is observed at impact parameters larger than the hump position and it is almost negligible for lower impacts, consistent with \cite {portocar98}.

In proton induced showers, the effect of the limited detector FOV is pronounced for all impact parameters and increases with the distance between the core axis and the detector position. The light absorption in the atmosphere reduced the mean densities by the same factor for all distances from the core axis independent of the primary particle. 

Figures 7c and 7d show the comparison between the lateral distributions calculated for two different observational heights - 2.2 km (MC set III) and 4 km (MC set IV) above sea level. The dashed and solid lines in plots correspond to 4.4 km and 2.2 km respectively. 
The hump observed in $\gamma$-rays  (figure 7c) is closer to the core axis for observation levels of 4 km due to a geometrical effect - the distance between the shower maximum and observation altitude is smaller. At this level, also the density decreases faster for impact parameters beyond the hump position, while close to the core axis higher densities are observed for the same primary energy, as shown in \cite{aha97,portocar98}. This effect was expected, since the average shower size is almost the same on the both simulated observation levels.\\
The same dependence on the observation altitude occurs in a proton shower. The average density close to the core axis is higher at 4.4 km than at 2.2 km and a faster decrease with the impact parameter has been obtained for the lower simulated observation depth.

\begin{figure}
\begin{center}
\includegraphics*[width=14cm]{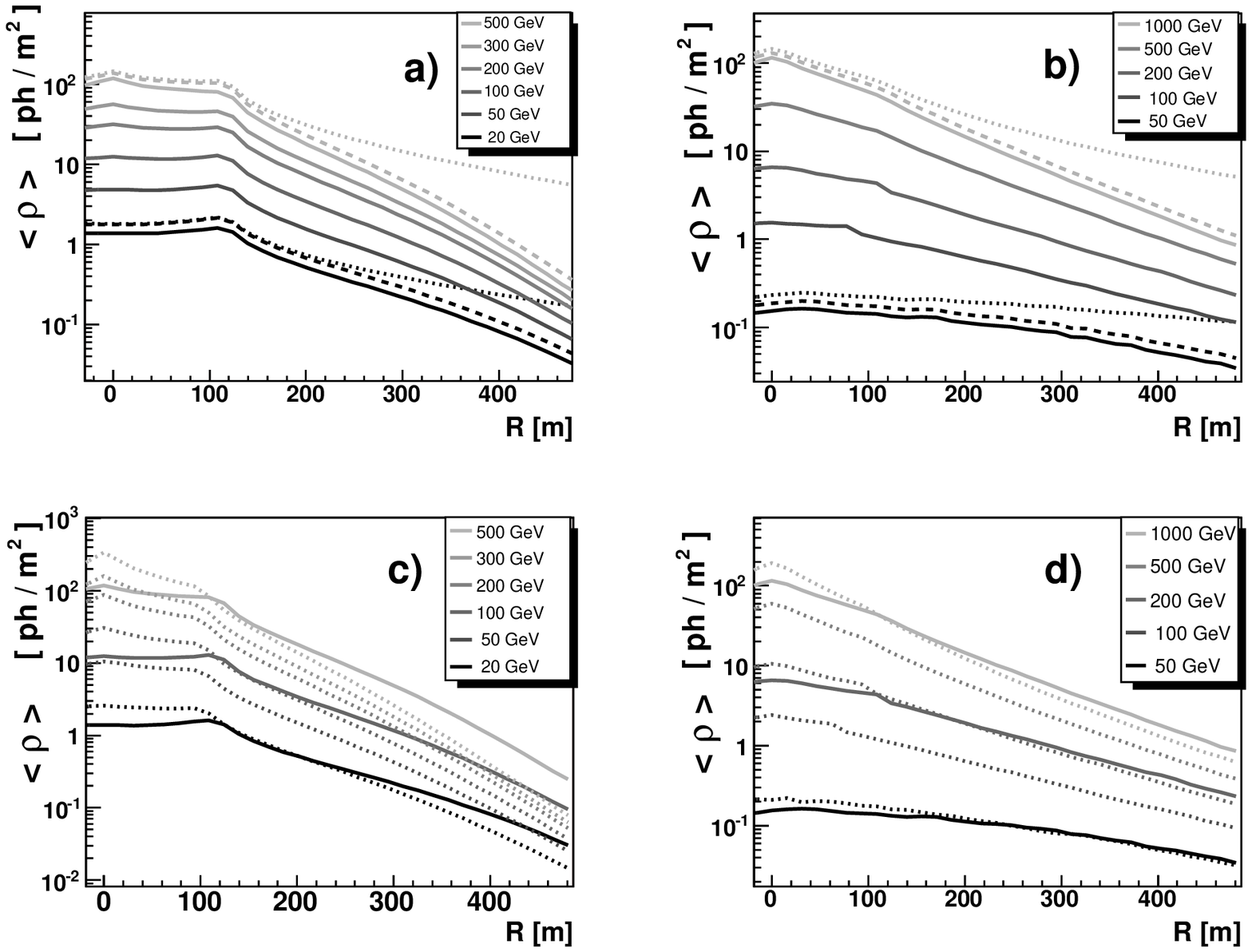}
\end{center}
\caption{The average lateral density distribution for all simulated primary energies: {\bf a)} primary $\gamma$-rays at 2.2 km a.s.l. {\bf b)} primary protons at 2.2 km a.s.l. Dotted, dashed and solid lines present results of MC set I, II and III respectively.\\
Comparison between lateral density distributions obtained at observations levels of 2.2 km (solid lines - MC set III) and 4.0 km (dotted lines - MC set IV): {\bf c)} for primary $\gamma$-rays (at 2.2 km - energies of 20, 100 and 500 GeV) ; {\bf d)} for primary protons (at 2.2 km - energies of 50, 200 and 1000 GeV.}

\label{later}
\end{figure} 

Figure 8a shows relative fluctuations of the density ($RF_{\rho}$) calculated from the same MC as in figure 7a for $\gamma$ primaries. The relative dispersion of the density diminishes with increasing primary energy (see solid lines in the plot - MC set III). This fluctuations are large at small impact parameters. It is shown in section 3.4 that Cherenkov photons close to the core axis are correlated with the production depths near to observation level, where the relative flutuations of the number of  the produced photons are very large (see Figure 3a).    
The local maximum of the presented in Figure 8 ratio pronounces the hump position for primary energies below or equal to 200 GeV. This feature disappears for higher energies. In the case of all produced photons (dotted lines), the ratio of the RMS deviation of the density to the mean is constant above the hump at a fixed energy. The inclusion of the limited telescope FOV in the simulations (dashed lines) enlarges the relative fluctuations significantly for impact parameters above 200 m. The effect of light absorption does not influence  the presented ratio. In the case of the MC set III, the relative fluctuations of the density do not differ for energies above 100 GeV and impact parameters above 250 m. It has been shown in \cite{sinha95} that for low energy $\gamma$'s, the relative dispersions of the density are almost stable below the hump for a detector size of 1 $m^2$. This is not true for a detector size of 240 $m^2$. My simulations show that the relative dispersion of the density decreases as the impact parameter increases from 0 to the hump position for 240 $m^2$ detectors (see figures 8a and 8c). 

The ratio of the RMS deviation of the Cherenkov photon density to its mean has been calculated for primary protons and is shown in figure 8b. The solid lines represent the results obtained from MC set III. The relative fluctuations decrease with the primary energy in the whole simulated impact parameter range. The ratio calculated for primary protons is higher than for primary $\gamma$-rays with the same shower size (the primary energy of the $\gamma$-ray is around three times lower than the primary energy of the proton). Additionally, the ratios obtained from all produced photons (dotted lines) and from photons in the detector FOV (dashed lines) are plotted in the figure for primary energies of 50 GeV and 1 TeV. Both effects enlarge the relative fluctuations at the lowest simulated primary energy, while only the limited FOV causes an increase of the ratio at the highest energy for impact parameters above 150 m. Large fluctuations in the curves at a primary energy of 50 GeV may indicate that the number of simulated events was too small. On the other hand all other curves (obtained for this energy and particle) are rather smooth. Relative fuctuactions for proton induced showers at a primary energy of 50 GeV are very high and it has been plotted to present the level of the $RF\rho$, not the exact number. 

The presented ratio was obtained for two different observation levels, which is presented in figure 8c and 8d for primary photons and protons, respectively. The shapes of the curves change a little bit. The relative dispersions are smaller at a height of 4 km (dotted lines) than at 2.2 km (solid lines) only at distances between 30 m and 100 m from the shower axis at $\gamma$-ray energies above 100 GeV. For the rest of the simulated energies and impact parameters, a lower ratio is expected for an altitude of 2.2 km a.s.l. for photon primaries. The average photon density expected at 4.4 km is lower than at 2.2 km for distances above 100 m which is the reason for larger relative fluctuations within this range. The explanation of changes close to the telescope axis is different. At a low impact parameter, the Cherenkov light is produced close to the observation level (see section 3.4), and at a height of 4.4 km one may expect some electrons and positrons even in a low energy electromagnetic cascade. The mean number of the charged particles is small and their statistical fluctuations are relatively large, so the expected fluctuations of the photon density are also relatively high (higher at 4.4 than at 2.2 km).
The relative fluctuations of the density are higher at an observation level of 4 km (dotted lines) than at 2.2 km (solid lines) at impact parameters above 100 m and very close to the shower axis for primary protons (figure 8d).

It has been checked that the number of photons within each detector does not have a Poissonian distribution in all MC sets, regardless of the primary particle type. The photons registered by one detector are not independent. Similar results were presented in \cite{bhat98} where the calculations were done for a detector size of 4.45 $m^2$. However, an approximately Poissonian distribution was shown in \cite{sinha95} for a 1 $m^2$ detector located farther away than the hump position (of primary $\gamma$-rays). Here one would expect the Poissonian fluctuations to be larger relative to the intrinsic fluctuations in the shower development due to the much smaller mirror area.

\begin{figure}
\begin{center}
\includegraphics*[width=14cm]{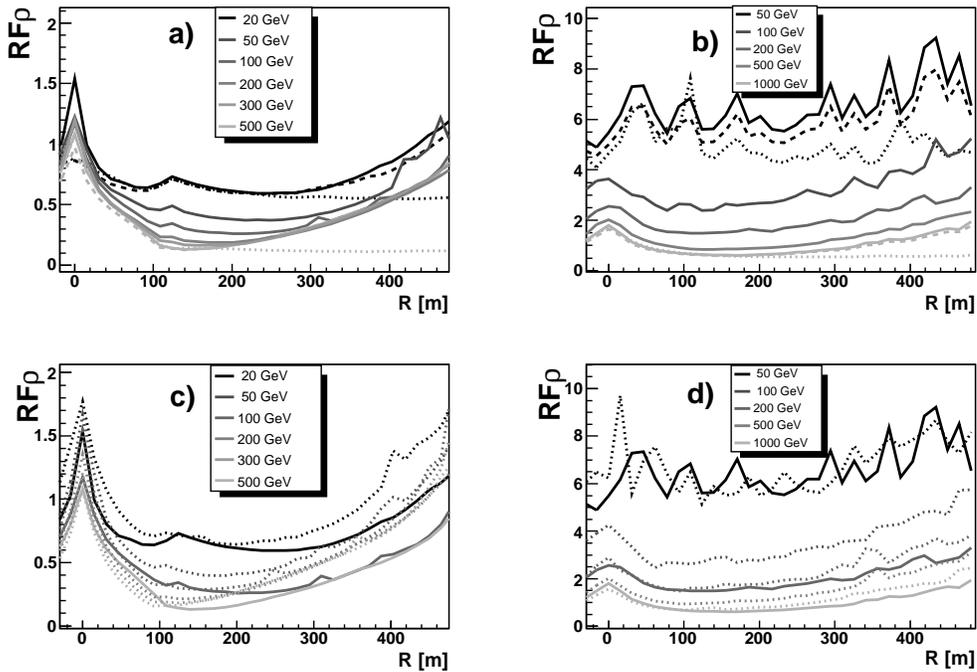}
\end{center}
\caption{The relative dispersions of the Cherenkov light density on the ground for all simulated primary energies: {\bf a)} primary $\gamma$-rays at an altitude of 2.2 km a.s.l.- dotted, dashed and solid lines correspond to the MC sets: I,II and III respectively; {\bf b)} the same as a) but for primary protons; {\bf c)} comparison between observations levels 2.2 (solid lines) and 4.0 km (dotted lines) - primary $\gamma$-rays (at 2.2 km - energies of 20, 100 and 500 GeV); {\bf d)} the same as c) but for primary protons (at 2.2 km - energies of 50, 200 and 1000 GeV).}

\label{later_rel}
\end{figure} 

The relative dispersions of the density have been calculated for detectors, which are two and four times larger than 240.25 $m^2$, and the results are shown in figure 9a for primary $\gamma$-rays. The ratio of the RMS deviation of the density to the mean density is almost independent on the detector size for all simulated energies and impacts parameters larger than 30 m. However, very close to the shower axis, lower relative fluctuations are obtained for larger detectors. 

In IACT experiments, the $\gamma$/hadron separation is efficient for impact parameters larger than around 40 m. The SIZE of the measured image is a sum of the signal from PMT's surviving the so called cleaning procedure. The SIZE is the observable which dependens on the Cherenkov photon density - in first order approximation it is proportional to the density. If enlarging the detector area does not change the relative dispersion of the density, then the relative fluctuations of the SIZE are not sensitive to the detector area (in the investigated area range: 240 - 960 $m^2$). A very significant improvement in the energy resolution of the primary particle with enlarged reflector size cannot be expected because the primary energy estimation is based on the SIZE and DIST parameters (DIST is the distance between the camera and image centres). This conclusion applies to measurements using one telescope only. One should expect that the real fluctuations of the measured light are larger due to the NSB (which is not taken into account in the presented simulations). The fluctuations of the number of photons (from NSB) are Poissonian and mirror size dependent. The influence of the NSB on the quality of the measurement is larger at lower primary energies. The estimations of the energy resolution shown in this paper are the limits obtained only for intrinsic fluctuations of the shower. The NSB is an additional reason for the difficulties with the low energy event measurements. When the trigger threshold is too low most of the recorded events are due to accidental triggers. This causes problems with too high tigger rates and efficient $\gamma$-ray selection.\\ 
The registration of the same event by more telescopes gives the possibility of a better energy resolution \cite{hoffman00}. The stereoscopic technique allows to estimate the shower maximum altitude, which results in better impact parameter determination and finally in better energy resolution, such as $9\%$ - $12\%$ \cite{hoffman00}. The energy resolution in the MAGIC experiment has been estimated as $20\%$ - $40\%$ \cite{unfol} which is consistent with the relative dispersions of the density shown in figure 9a for distances from 60 m to 200 m and energies above 50 GeV. The accuracy of the energy determination has been estimated as $20\%$ in the HEGRA experiment \cite{aha99} while VERITAS has shown an energy resolution of $10\%$- $20\%$ \cite{hanna08}. 
The estimation of the energy resolution obtained in this analysys for an ideal single IACT is presented in Table 3 for two investigated observation levels. Energy resolutions are better on altitudes of 4 km a.s.l. for primary energies above 200 GeV at impact parameters between 40 m and 100 m only. For larger impacts the expected  density of the light is lower at 4 km altitude (mostly due to the geometrical effect) what results in higher relative $RF\rho$. Well above 200 GeV, at very high altitude  the shower development will be truncated at the observation level and this will have a very significant negative impact on the energy resolution. The same effect causes an increase of the $RF\rho$ for low impact parameters (below 40 m) at the observation level of 4 km in comparison to 2.2 km. The lower the primary energy of the $\gamma$-ray, the worse the energy resolution that is expected at the higher simulated observation level.    

\begin{table}
\caption {\label{tab3}Estimation of the energy resolution for an ideal single Cherenkov telescope (mirror area of 240.25 $m^2$)}
\begin{indented}
\item[]\begin{tabular}{@{}*{5}{l}}
\br
MC set & primary energy  & impact &parameter & \cr
 & in GeV  & 40 m & 60 m & lowest value\cr
\mr
III &  20  & 0.75 & 0.70 & 0.60 \cr
III &  50  & 0.65 & 0.58 & 0.38 \cr
III & 100  & 0.62 & 0.50 & 0.27 \cr
III & 200  & 0.55 & 0.44 & 0.19 \cr
III & 300  & 0.55 & 0.43 & 0.17 \cr
III & 500  & 0.55 & 0.41 & 0.14 \cr
\mr
IV &  20  & 0.95 & 0.80 & 0.65 \cr
IV &  50  & 0.75 & 0.60 & 0.40 \cr
IV & 100  & 0.64 & 0.50 & 0.30 \cr
IV & 200  & 0.54 & 0.40 & 0.21 \cr
IV & 300  & 0.48 & 0.35 & 0.19 \cr
IV & 500  & 0.43 & 0.30 & 0.16 \cr

\br
\end{tabular}
\end{indented}
\end{table}
   
Figure 9b shows a comparison between the relative dispersions of the Cherenkov photon density calculated for different detector sizes for proton primaries. With increasing detector size, the lower ratio obtained for the low energy proton. The differences decrease as the energy increases, and practically disappear at a primary energy of 500 GeV.

\begin{figure}
\begin{center}
\includegraphics*[width=14cm]{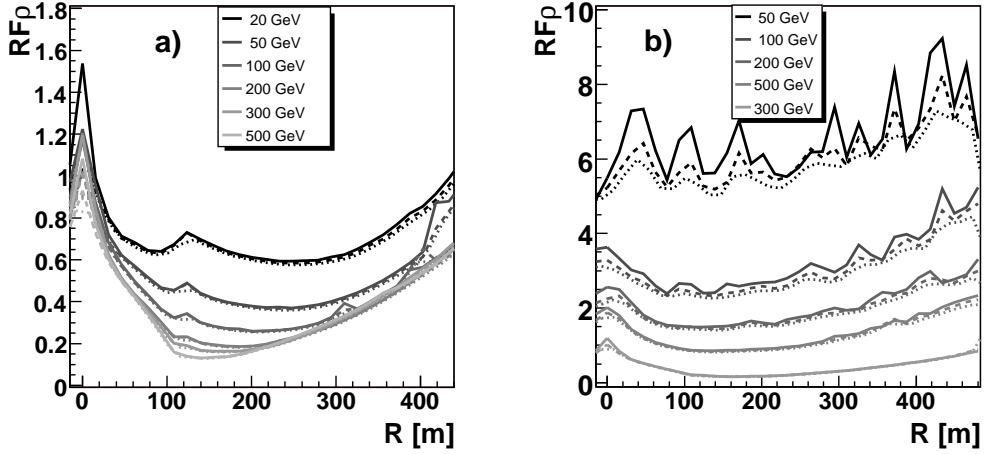}
\end{center}
\caption{The relative dispersions of the Cherenkov light density obtained for different detector sizes: 240, 480 and 960 $m^2$ from MC set III (solid, dashed and dotted lines): {\bf a)} primary $\gamma$; {\bf b)} primary proton.}

\label{later_rel124}
\end{figure} 

\subsection{Correlations between longitudinal and lateral distributions}

The Cherenkov light produced in the atmosphere hits the ground at the distance R from the core axis which depends on the zenith angle of the photon ($\phi$) and the height of its production (h). The simple formula R = h * tan($\phi$) describes this dependence. If one assumes that all photons are produced on the shower axis and their zenith angles are the Cherenkov angles of ultrarelativistic electrons, then R is limited by $R_{max}$. $R_{max}$ is independent of the primary energy. It has been checked that $R_{max}$ is around 130 m for the observation level of 2.2 km a.s.l. Cherenkov photons in the $\gamma$ cascade hit the ground farther than $R_{max}$ because charged particles are not produced on the shower axis and they do not travel along this axis. The angle between the velocity of the electron (or positron) and the direction of the primary $\gamma$-ray originates from the electromagnetic processes during the cascade development (like bremstrahlung and pair production) and the multiple Coulomb scattering effect of $e^+$ (or $e^-$).

The correlation coefficients (denoted in figures as Corr. coeff.) between the number of the produced photons at a fixed level and the density of light at a chosen distance from the core (R) are shown in figures 10 and 11 for primary $\gamma$-rays and protons respectively. Four depths have been chosen to present the correlation coefficients before (250 $g/cm^2$), close to the mean shower maximum of the high energy $\gamma$ cascade (350 $g/cm^2$ and 450 $g/cm^2$) and near to the observation level (750 $g/cm^2$). Adequate heights (from the US standard atmosphere model) are: 10.5, 8.1, 6.5 and 2.6 km a.s.l. The solid, dashed and dotted lines show the results of  MC sets: III, II and I respectively. The large and positive value of the coefficient in the figures indicates the depth where most of the photons hitting the detector at distance R were produced.
The correlations are more pronounced at a higher energy than at a lower energy of the primary $\gamma$-ray. The Cherenkov light seen by the telescope (MC set III) close to the core axis originates mainly near the observation level (figure 10d). The distance between the photon and core axis increases as the production height increases. The light produced near the shower maximum is the main contribution to the density at the hump position (figure 10b). The inclusion of the limited detector FOV in the simulations changes the calculated correlation coefficients significantly for all atmospheric depths. This effect favours photons coming from corresponding heights (compare the dotted and the dashed lines). The atmospheric absorption effect has a negligible impact on the calculated coefficients. 

For primary protons, the correlations between the longitudinal and lateral distributions are much less visible (figure 11). It seems that in the lowest simulated energy of 50 GeV, the Cherenkov light detected at a fixed distance from the core is not connected with any favourite production altitude. The correlations are more pronounced when the primary energy of the proton increases, and show a similar tendency as $\gamma$ cascades - the photons seen by telescope which are farther from the shower axis come from lower atmospheric depths.
\\

\begin{figure}
\begin{center}
\includegraphics*[width=14cm]{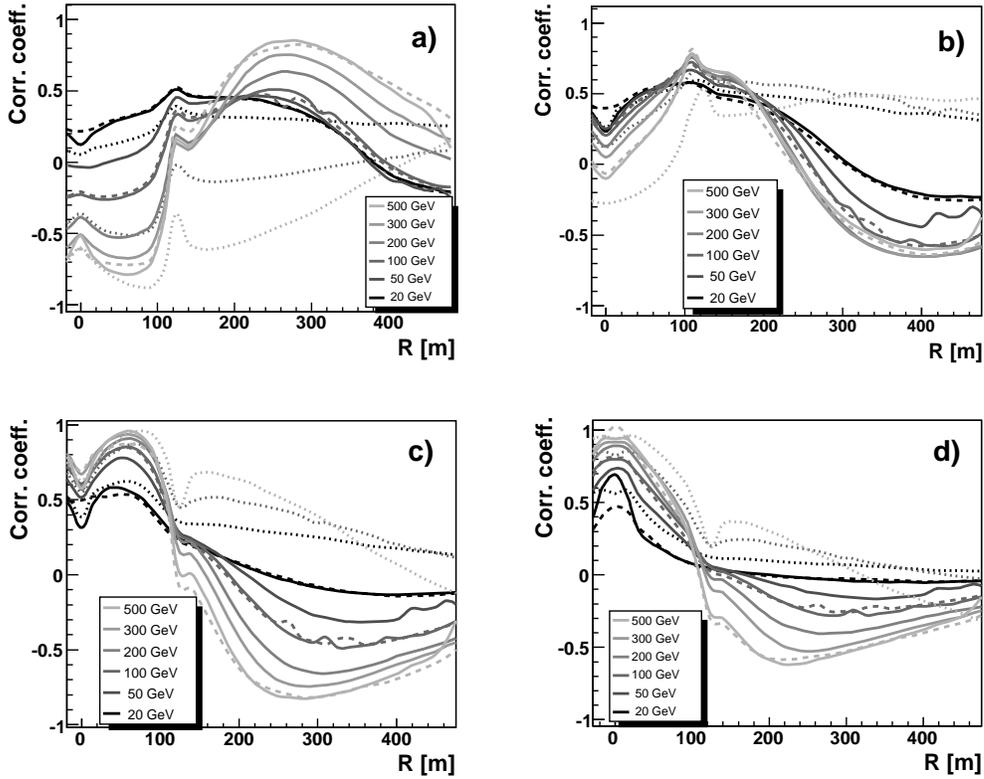}
\end{center}
\caption{The correlations between longitudinal and lateral distributions for primary $\gamma$-rays at depths: {\bf a)} 250 $g/cm^2$; {\bf b)} 350 $g/cm^2$; {\bf c)} 450 $g/cm^2$; {\bf d)} 750 $g/cm^2$. Dotted, dashed and solid lines correspond to to all produced photons (MC set I), photons hitting the telescope mirror within the 5 deg. field of view, neglecting atmospheric absorption (MC set II) and photons incident on the telescope mirror taking into accout inefficiency from atmospheric absorption and limited FOV (MC set III), respectively. The results of the MC sets I and II are plotted for energies of 20, 100 and 500 GeV only.}

\label{corr_gamma}
\end{figure} 
\begin{figure}
\begin{center}
\includegraphics*[width=14cm]{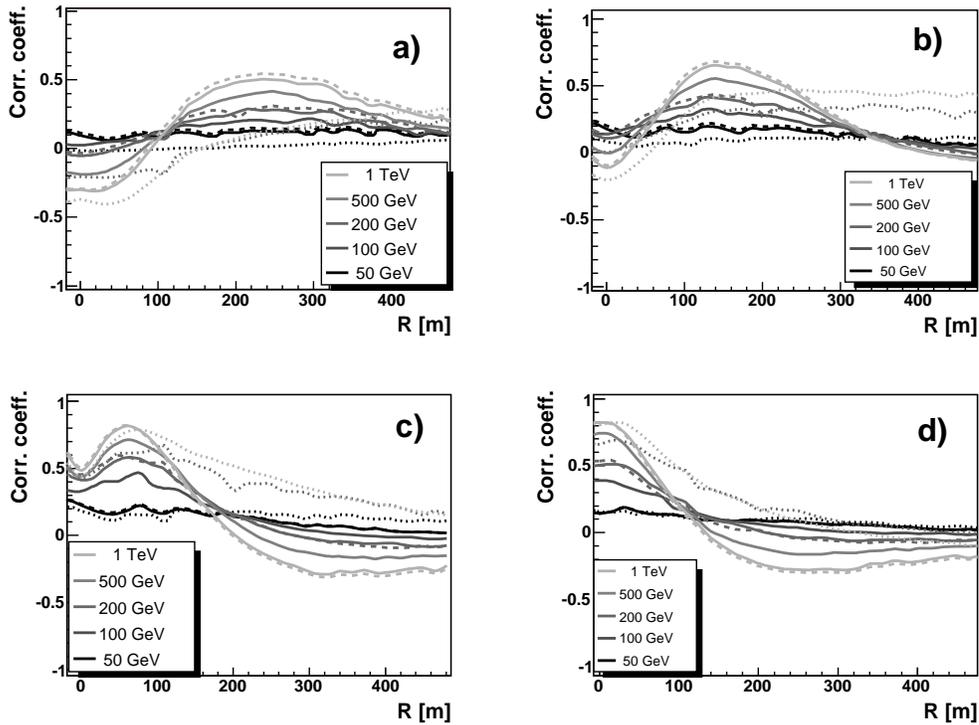}
\end{center}
\caption{The correlations between longitudinal and lateral distributions for primary protons at depths: {\bf a)} 250 $g/cm^2$; {\bf b)} 350 $g/cm^2$; {\bf c)} 450 $g/cm^2$; {\bf d)} 750 $g/cm^2$.  Dotted, dashed and solid lines correspond to to all produced photons (MC set I), photons hitting the telescope mirror within the 5 deg. field of view, neglecting atmospheric absorption (MC set II) and photons incident on the telescope mirror taking into accout inefficiency from atmospheric absorption and limited FOV (MC set III), respectively. The results of the MC sets I and II are plotted for energies of 50, 200 and 1000 GeV only.}

\label{corr_proton}
\end{figure} 

\section {Conclusions}

There are differences in the longitudinal and lateral distributions obtained from differnt MC simulations sets. The relative fluctuations of all studied observables decrease as the primary energy increases for both types of simulated particles. The low energy showers show larger intrinsic fluctuations than high energy EAS, even as the mirror area is increased and Poissonian fluctuations are reduced.

The limited detector FOV causes a shift of the caculated average shower maximum towards lower depths, while the effect of absorption is opposite. As a result, the shower maximum obtained from MC set III is 20 $g/cm^2$ higher than that calculated from all Cherenkov photons (MC set I). This shift is negligible for protons below 500 GeV. 

For $\gamma$-rays the relative fluctuations of the number of Cherenkov photons produced at a fixed level increase below the shower maximum if we calculate them from the Cherenkov light in the limited FOV. This ratio does not change above the shower  maximum height. The same relative fluctuations calculated for proton induced showers are more sensitive to the inclusion of the limited telescope FOV at lower than at higher primary energies. These relative dispersions do not depend on the absorption of Cherenkov light in the atmosphere for both primary particles at all simulated energies.

As expected the total number of photons in the electromagnetic cascade is larger if the cascade starts deeper in the atmosphere, while there are no correlations between the shower size and the first interaction height in hadronic showers. The mean shower size is a linear function of the primary energy in all MC simulations sets. The same fraction of mean shower size does not fulfil the detector conditions for primary $\gamma$-rays above 100 GeV - below this limit the part of all photons which survive the detector criteria increases slightly with the primary energy. More Cherenkov photons fulfil the detector conditions at higher than at lower primary proton energies. The relative fluctuations of the showers size are significantly larger below 100 GeV and 300 GeV for primary photons and protons, respectively.

The lateral density distributions calculated from photons in the limited FOV is significanty lower than the one obtained from all photons at higher distances from the core axis, while at low impact parameters most of the light is in the telescope FOV. The density decreases by the same factor at all investigated distances due to the absorption of the light.
The fluctuations of the Cherenkov light density are not Poissonian, but rather are dominated by intrinsic shower fluctuations when one considers large mirror areas. The relative dispersions of the density are much more pronounced in proton than $\gamma$-ray showers. In both cases, they are higher for Cherenkov photons in a detector with limited FOV than for a detector that can collect all produced photons. The absorption in the atmosphere has no influence on these fluctuations.
For $\gamma$-rays the relative fluctuations of the Cherenkov light density on the ground are quite large at small impact parameters.

The relative density fluctuations are independent of the detector size (in the investigated range - from 240 $m^2$ up to 960 $m^2$) for $\gamma$-rays beyond 40 m distance. In the case of protonic showers, a similar independence is observed for energies above 500 GeV. Supposing that the ratio between the RMS deviation of the density to its mean is an estimation of the primary energy resolution, even a four times larger reflector surface than that of the MAGIC telescope, a single telescope cannot achieve a better energy resolution than presented in this paper.
The simulations show that better energy resolution in a single IACT can be achieved by building the experiment at an altitude of 4 km a.s.l., but this only concerns primary energies above 200 GeV at impact parameters between 40 and 100 m. At lower energies the energy resolution should be better on an observation level of 2.2 km.
The energy resolution of a single IACT is limited by the fluctuations in the shower development itself which causes the difficulties of the detection of low energy $\gamma$-rays.

\ack
This work was supported by the Polish KBN grant No. N N203390834.


\section*{References}

\begin{thebibliography}{10}

\bibitem{whipple} {Weekes T C {\it et al} 1989 {\it Astrophys. J.} {\bf{342}} 379}
\bibitem{hillas} {Hillas A M 1985 {\it Proc. 19th Int. Cosmic Ray Conf. (La Jolla)} vol 3, p 445 (1985)} 
\bibitem{aha97} {Aharonian F {\it et al} 1997 {\it Astropart. Phys.},{\bf{6}} 343 } 
\bibitem{portocar98}{ Portocarrero C E and Arqueros F  198 {\it J.Phys.G: Nucl. Part. Phys.} {\bf{24}} 235}
\bibitem{aha99} {Aharonian F {\it et al} 1999 {\it Astronomy and Astrophysics} {\bf 342} 69} 
\bibitem{hoffman00} {Hoffman W {\it et al} 2000 {\it Astropart. Phys.} {\bf 12} 207} 
\bibitem{unfol} {Albert J {\it et al} 2007 {\it Nucl. Instrum. Methods Phys. Res.} A {\bf 583} 494}
\bibitem{hanna08} {Hanna D {\it et al} 2008 {\it Nucl. Instrum. Methods Phys. Res.} A {\bf 588} 26}
\bibitem{bhat98} {Chitnis V R and Bhat P N 1998 {\it Astropart. Phys.} {\bf 9} 45} 
\bibitem{sinha95}{ Sinha S 1995 {\it J.Phys.G: Nucl. Part. Phys.} {\bf{21}} 473}
\bibitem{bhat02} {Chitnis V R and Bhat P N 2002 {\it Experimental Astronomy} {\bf{13}} 77}
\bibitem{contreras06} {Contreras J L {\it et al} 2006 {\it Astropart. Phys.},{\bf{26}} 50 } 
\bibitem{bario} {Barrio J A {\it et al} 1998 The MAGIC Telescope Design Study (Munich)}
\bibitem{bax04} {Baixeras C {\it et al} 2004 {\it Nucl.Instrum. Methods Phys. Res.} A {\bf 518}  188}
\bibitem{albert} {Albert J {\it et al} 2005 {\it Astropart Phys} {\bf 23} 493}
\bibitem{heck} {Heck D {\it et al} 1998 Technical Report FZKA 6019 (Forschungszentrum Karlsruhe)} 
\bibitem{knapp} {Knapp J and Heck D 2004 EAS Simulation with CORSIKA: A User´s Manual} 
\bibitem{sokol} {Sokolsky P 1989 ``Introduction to Ultrahigh Energy 
Cosmic Ray Physics'' Addison-Wesley}
\bibitem{raosinha95}{ Rao M V S and Sinha S 1988 {\it J.Phys.G: Nucl. Part. Phys.} {\bf{14}} 811}
\bibitem{porter73}{Porter N A 1973 {\it Nuovo Cimento Lett.} {\bf 8} 481}
\bibitem{bowden92}{ Bowden C C G {\it et al} 1992 {\it J.Phys.G: Nucl. Part. Phys.} {\bf{18}} L55}
\bibitem{lang94}{ Lang M J {\it et al} 1994 {\it J.Phys.G: Nucl. Part. Phys.} {\bf{20}} 1842}
\bibitem{comm08} {Commichau {\it et al} 2008 {\it Nucl.Instrum. Methods Phys. Res.} A {\bf 592}  572}




\end{thebibliography}
\end{document}